\documentclass[12pt]{article}
\usepackage{graphics}
\input epsf
\usepackage{graphics}
\usepackage[dvips]{color}
\usepackage{multicol}
\usepackage{amsmath,color,fancybox,graphics,graphpap,rotating}
\definecolor{white}{rgb}{1,1,1}
\definecolor{yellow}{rgb}{0.95,0.75,0.1}
\definecolor{red}{rgb}{0.5,0,0}
\definecolor{green}{rgb}{0,1,0}
\definecolor{blue}{rgb}{0,0.5,1}
\definecolor{bgcolor}{rgb}{0.94,0.91,0.78}%still darker
\definecolor{lblue}{rgb}{0,0.8,1}
\definecolor{dblue}{rgb}{0,0,.6}
\definecolor{dgreen}{rgb}{0,0.3,0}
\definecolor{lila}{rgb}{0.8,0,0.8}
\definecolor{violet}{rgb}{1,0,1}
\definecolor{grey}{rgb}{0.3,0.3,0.3}
\definecolor{turquoise}{rgb}{0,.9608,1}

\definecolor{contoura}{rgb}{0,0,1}
\definecolor{contourb}{rgb}{0,1,1}
\definecolor{contourc}{rgb}{0,1,0}
\definecolor{contourd}{rgb}{0.95,0.75,0.1}
\definecolor{contoure}{rgb}{1,0,0}
\definecolor{contourf}{rgb}{1,0,1}

\def\lsim{\raise0.3ex\hbox{$\;<$\kern-0.75em\raise-1.1ex\hbox{$\sim\;$}}}
\def\gsim{\raise0.3ex\hbox{$\;>$\kern-0.75em\raise-1.1ex\hbox{$\sim\;$}}}

\newcommand{\nee}{\nonumber \end{eqnarray}}

\newcommand{\be}{\begin{eqnarray}}
\newcommand{\ben}{\begin{eqnarray}\nonumber}
\newcommand{\ee}{\end{eqnarray}}

\begin{document}

\title{Finite Matter Resolution of the Cosmological Entropy Problem}
\author{
L. Clavelli\footnote{Louis.Clavelli@Tufts.edu $\,,\,$lclavell@bama.ua.edu}\\
Dept. of Physics and Astronomy, Tufts University, Medford MA 02155\\
Dept. of Physics and Astronomy, Univ. of Alabama, Tuscaloosa AL 35487}
\date{}
%\date{\today}
\maketitle

\begin{abstract}

We discuss how entropy bounds, which are not respected in the standard cosmology, constrain the parameters of a 
previously suggested cosmology with a finite total mass \cite{CG1},\cite{CG2}. 
In that alternative cosmology the matter density was postulated to be a spatial delta function at the time of the big bang thereafter diffusing rapidly outward with constant total mass. 
Also discussed here are some related issues including the cosmic onion question, the information content of the universe, and the question of whether light trapping regions exist on a cosmic scale.
%\pacs{PACS numbers: 11.30.Pb, 12.60.J, 13.85.-t}
\end{abstract}

keywords: Cosmology, Bekenstein Entropy bound, Finite Mass

\section{Introduction}

The Bekenstein \cite{Bekenstein} bound on entropy can be formulated as stating that no object of radius $r$ can have an entropy greater than the entropy of a black hole of that Schwarzschild radius.  
The bound was originally derived in the Gedenkenexperiment in which a
body of given entropy is lowered into a black hole of slightly greater mass.
It can, perhaps, be most easily seen by noting that any object that can collapse into a black hole due to gravity or any other force must begin with an entropy less than that of the black hole since the entropy in any process should increase or remain constant.
The standard homogeneous and isotropic cosmology violates the Bekenstein bound since the entropy in a sphere of radius $r$ grows
as $r^3$ inevitably exceeding the entropy of a black hole of that
horizon radius which grows as $r^2$.  Thus this cosmology, although consistent with Einstein's equations, is inconsistent with thermodynamics if
a dark energy transition to a big-crunch with negative cosmological constant is physically possible and, in any case, is inconsistent
with the Bekenstein bound.
 
The entropy of a black hole of Schwarzschild radius $r$ satisfies
\be
     \frac{S_{BH}(r)}{2 \pi k} =  \frac{r^2}{2 {L_{Pl}}^2}
\label{Schwarzschild}
\ee
where the Planck length is
\be
     L_{Pl} = \sqrt{\hbar G/c^3} = 1.6\cdot 10^{-35}\,\displaystyle{m}.
\ee

\section{Bekenstein bound in a finite mass universe}
If the entropy of any object of radius $r$ is $S(r)$, the Bekenstein bound requires
\be
     S(r) < S_{BH}(r)
\label{BekensteinBound}
\ee

Known astrophysical objects including main sequence stars, white dwarfs, and neutron stars easily satisfy this bound
but the standard homogeneous cosmology, inevitably violates the Bekenstein bound for large enough $r$ as stated above.  Various attempts to deal with
the problem within the context of the infinite homogeneous universe have been summarized in ref. \cite{MasoumiMathur}.

The entropy density of an object of density $\rho$, pressure $p$, and chemical potential $\mu$ at temperature $T$ satisfies \cite{OlivePeacock}

\be
    \frac{s(r)}{2 \pi k} =  \frac{1}{2 \pi k T}\, (\rho(r) + p(r) - \mu\, n(r)) 
\label{entropydensity}
\ee
where $\rho$ is the energy density, $p(r)$ is the pressure, and $n(r)$ is the number density of constituents. The chemical potential $\mu$ is
a mass factor for the constituents.  Clearly, if the universe is infinite and homogeneous and the entropy density follows this equation, the total entropy in a sphere of radius $r$ grows as $r^3$ violating the Bekenstein bound. This has motivated several alternative entropy bounds, the holographic entropy bound \cite{tHooft},\cite{Susskind}, a covariant entropy bound \cite{Bousso}, and a causal entropy bound \cite{BrusteinVeneziano}. All of these alternative bounds involve heuristic and unexplained assumptions.  In addition it has been proposed \cite{Frampton} that
the entropy problem could be avoided in a cyclic cosmology if the entropy is
"reset" to zero at each turnaround and each bounce.  

Dark energy does not contribute to the entropy since $\mu=0$ and $p = - \rho$. 
%...
Observations are consistent with this latter equality to within $10\%$.  There is no clear way to solve the entropy puzzle by assuming the violation of this equality and associating an entropy to dark energy.  In any case, in this paper we
restrict our considerations to the entropy constraints on matter.
%...  
For a photon gas the chemical potential also vanishes and the pressure density is $\rho /3$ so that
\be
   \frac{s_\gamma(r)}{2 \pi k} =  \frac{2}{3 \pi}\,\frac{1}{k\,T}\rho(r) =  \frac{2}{3 \pi k T}\,<E_\gamma>\,n_\gamma(r) \quad .
\ee
Over a wide range of temperatures the number density of photons is known to be proportional to the number density of baryons:
\be
    n_\gamma (r) \approx \frac{10^{10}}{6.05} n_b (r) \quad .
\ee

The average photon energy in a black body of photons is
\be
     <E_\gamma> \approx 2.701 \cdot k\,T \quad .
\ee
If we integrate up to radius $r$, the entropy, $S(r)$, in a sphere of that radius is therefore related to the total number of baryons, $N_b(r)$, in that sphere.
\be
    \frac{S_\gamma (r)}{2 \pi k} \approx \frac{5.40}{3 \pi}\cdot \frac{10^{10}}{6.05} N_b(r)\quad .
\label{SgammaOfr}
\ee

The large photon to baryon ratio implies that we can safely neglect
the entropy carried by baryonic matter.
In the standard homogeneous cosmology $N_b(r)$ grows as $r^3$ so that
the total entropy from eq.\,\ref{SgammaOfr} inevitably exceeds the entropy of a black hole of that Schwarzschild radius as alluded to above.

Clearly, a possible resolution is the construction of a model where the number of baryons in a sphere of radius $r$ saturates or grows no more than quadratically with $r$. 
%...
An interesting potential solution \cite{Pietronero}\cite{deVega} of the entropy problem which has now 
apparently been ruled out  is the possibility of a fractal universe with dimension
$d_H \le 2$.  In a fractal universe the mass and entropy inside a sphere of radius $r$  increase with $r$ as $r^{d_H}$.  
It is found instead that, although at small distances the universe shows fractal structure
of dimension $\le 2$, beyond $100 \displaystyle{Mpc/h}$ and out to near Hubble scales the matter density smooths out and becomes consistent with the 
Friedmann, Robertson, Walker (FRW) universe \cite{Hogg}\cite{Tegmark}.
Although one could seek ways to avoid the counter-indications to a fractal
universe,  we pursue in this paper the simpler possibility that the universe
contains a finite mass with inhomogeneity setting in at the Hubble scale or above.
%...
Two such models were proposed in ref.\cite{CG1,CG2}.  

In the first, a matter density with infinite range but finite total mass took the form
\be
    \rho(r,t) = \frac{M}{(\sqrt{\pi} R a(t))^3}\, e^{-r^2/(R a(t))^2}
\quad .
\label{rhom}
\ee
Here $R$ is the scale of inhomogeneity and $a(t)$ is approximately the
scale factor of the Friedmann-Robertson-Walker model since in regions of low curvature Hubble's law can be derived \cite{CG1}:
\be
    \vec{v} = \frac{\dot{a}}{a} \vec{r} = H \vec{r} \quad .
\ee 

$a(t)$ is taken to vanish at $t=0$, increase monotonically, and equal unity at present time.

In the second, a matter density with finite range, finite total mass, and a bubble topology took the form
\be
    \rho(r,t) = \frac{M}{5.15 R^3 a(t)^3}\, (3 -2 (r/(a(t)R)^2 )\, e^{-r^2/(R a(t))^2}\, \theta(3 -2 (r/(a(t)R)^2) \quad .
\label{bubble}
\ee
Both were obtained as the time-time component of the Einstein tensor for appropriately defined space-time metrics. 
In the present work we ignore possible 
effects of the off-diagonal components of the Einstein tensor taking the above densities as our starting point. 
The baryon number density is obtained in each case by dividing by the nucleon mass, $m_N$.  Clearly, if we integrate over a sphere of sufficiently large radius each model is consistent with the Bekenstein bound since the enclosed entropy asymptotes to a constant.  It could  however, be asked whether for smaller radius some constraint on the parameters $M$ or $R$ is obtained.
 
For simplicity we restrict our analysis here to the first model although the second model will not lead to qualitatively different results.

The analysis is most appropriately done in the co-moving frame defined by the co-moving coordinate
\be
    r_c = r/a(t) \quad .
\ee
In this frame the matter density takes the time independent form
\be 
    \rho_c (r_c) = \frac{M}{(R {\sqrt\pi})^3}\, e^{-{r_c}^2/R^2} \quad .
\label{rhoc}
\ee
As $R \rightarrow \infty$ with $M/R^3$ constant, the model approaches the standard homogeneous cosmology.
The number of baryons in a sphere of radius $r_c$ is
\be\nonumber
    N_b(r_c) &=& \frac{ M}{m_N (R \sqrt \pi)^3} \, \int_0^{r_c} 4 \pi {{r_c}^\prime}^2 d{r_c}^\prime \,e^{-{({r_c}^\prime /R)}^2}\\
  &=& \frac{2 M}{m_N {\sqrt \pi}}\,\gamma(3/2,{r_c}^2/R^2) 
\label{Nbofrc}
\ee 
where $\gamma$ is the incomplete $\Gamma$ function of the given arguments.

This can be expressed in terms of the co-moving matter density near $r=0$ from eq.\,\ref{rhoc} or the present density near $r=0$ in the expanding frame.
In the inhomogeneous universe of total mass M, 
\be
       M = \rho_c(0) R^3 \pi^{3/2}
\ee
where,
\be
     \rho_c(0) = \rho(0,{\displaystyle{now}}) = \Omega_b \rho_{critical}      = 0.24\; GeV/c^2/m^3 
      = 8.9 \cdot 10^{-45} \frac{M_N}{{L_{Pl}}^2 L_H} \quad .
\ee
Here we have found it convenient to express the co-moving density at the
origin in terms of the Planck and Hubble lengths.

Collecting factors it is seen that, in the co-moving frame, the entropy contained in a sphere of radius $r_c$ is specified by
\be
   \frac{S_\gamma (r_c)}{2 \pi k} \approx 5.3 \cdot 10^{-35} \frac{R^3}{L_H {L_{Pl}}^2}\, \gamma(3/2,{r_c}^2/R^2)
\ee
Relative to the entropy in a black hole of Schwarzschild radius $r_c$
from eq.\,\ref{Schwarzschild} this would be
\be
    \frac{ S_\gamma (r_c)}{S_{BH}(r_c)} \approx 1.06 \cdot 10^{-34} \frac{R}{L_H}\, \frac{R^2}{{r_c}^2} \gamma(3/2, {r_c}^2/R^2) \quad .
\label{EntropyRatio}
\ee
Consistency with the Bekenstein bound is obtained if this ratio is less than unity for all $r_c$.   

Although we have found it convenient to derive the constraint in the co-moving frame, the same constraint would be found in the frame of the accelerating universe with scale factor $a(t)$ where it would take the form
\be
       \frac{ S_\gamma (r)}{S_{BH}(r)}  \approx  1.06 \cdot 10^{-34} \frac{R}{L_H}\, \frac{R^2 a(t)^2}{{r }^2} \gamma(3/2, {r/(a(t) R)}^2) \quad .
\label{EntropyRatio2}
\ee
In either case, the function
\be
     \frac{1}{x^2} \gamma(3/2,x^2) = \frac{1}{x^2}\, \int_0^{x^2} dy
     \,\sqrt{y} e^{-y}
\ee
clearly vanishes at $x=0$ and $x=\infty$.  It never exceeds a value of
about $0.379$ which is obtained at $x \approx 0.968$.
Thus, consistency with the bound requires
\be
    0.40 \cdot 10^{-34} \frac{R}{L_H} < 1 
\label{RoLH}
\ee
or 
\be
     R < 3.4 \cdot 10^{60} m \quad .
\ee

The matter inhomogeneity implied by a finite value of $R$ is usually thought to be constrained by the isotropy of the Cosmic Background Radiation (CBR) and by the homogeneity of the type Ia Supernovae.
However, the CBR in general implies only that $R \ge 10^5\,L_H$ and, in 
the case that the Milky Way is near the $r=0$ position, a much smaller value of $R$ is possible.  The supernovae data do not extend much beyond redshift $z=1$ so they do not 
strongly constrain $R$.  In fact, indications from galaxy clusters
suggest \cite{Beckwith} a value \cite{CG2} of $R$ near the Hubble length, $L_H$.  If we treat the Beckwith result as a lower limit on 
the $R$ parameter, this and the Bekenstein bound imply the constraints
\be
     0.94 < \frac{R}{L_{H}} < 2.5\;10^{34}
\ee
The upper limit on $R$ from the entropy bound is not very restrictive but the 
fact that there is a finite upper limit at all is of interest.  Even this very weak upper limit allows the estimate that there is a negligible probability to find two identical unrelated humans in the universe. 
To see this assume there is no more than one human civilization per solar system with $10^{11}$ humans in each, $10^{11}$ solar systems
per galaxy and $10^{11}$ galaxies per Hubble length.  Then the number of humans in the finite matter universe is no more than $10^{33}\, (R/L_H)^3$.  Since this is much less than the number of distinct human genomes, $10^{3.6\,10^9}$, the probability of producing identical 
twins by random shuffling of genes is negligible.

The total information content in the universe is
\be
     I = \frac{S}{k \log 2}
\ee
This is infinite in the standard cosmology but finite and dependent on
$R$ in the finite matter cosmology.  This has a bearing on the ``cosmic onion" question in physics:  Are there a finite number of
constituent species and fundamental interactions or, in the onion analogy, is there no limit to the new fundamental physics that can be revealed as we peel off successive layers?  If the total information content in the universe is finite as in the finite matter model, there cannot be an infinite number of constituent species or fundamental gauge groups.  Thus, in this model, physics is finite at the fundamental level although there is still great scope for applied physics if and when the complexity of the fundamental laws is exhausted.

\section{Constraint from absence of light trapping}
A tighter bound on $R$ can be obtained if one postulates that the total matter inside any radius $r_c$ in the co-moving frame is less
than the Schwarzschild mass corresponding to that horizon length.
Then the universe will never form a large scale light trapping region.  The total matter inside $r_c$ is from eq.\,\ref{Nbofrc}
\be
   M_b(r_c)= m_N N_b(r_c) =  2 \pi R^3 \rho_c(0)\gamma(3/2,{r_c}^2/R^2)
\quad .
\ee
The mass of a black hole with horizon $r_c$ is
\be
     M_{BH}(r_c) = \frac{c^2}{2 G} r_c
\ee
so
\be\nonumber
   \frac{M_b(r_c)}{M_{BH}(r_c)} &\approx& 4 \pi \, \frac{G \rho_c(0)R^2}{c^2}\frac{R}{r_c} \gamma(3/2, (r_c/R)^2)\\ &=&0.073 \frac{R^2}{{L_H}^2}\frac{R}{r_c} \gamma(3/2, (r_c/R)^2)\quad .
\label{MoMBH}
\ee
The incomplete gamma function is such that for any $R/r_c$
\be
     \frac{R}{r_c} \gamma(3/2, (r_c/R)^2) < 0.466 \quad .
\ee
Thus, for any co-moving radius, $r_c$, the mass enclosed is guaranteed to be  less than a black hole of that horizon size if
\be
     R < 5.4 \,L_H \quad .
\label{NoTrapping}
\ee
Combining this no-cosmic-light-trapping requirement with the result of \cite{Beckwith} leads to tight upper and lower bounds on the inhomogeneity of the universe
\be
     0.94 \,L_H < R < 5.4 \,L_H \quad .
\ee 

If, on the other hand, the inhomogeneity found in ref. \cite{Beckwith} is not confirmed and  
\be
      5.4\,L_H < R < 2.5 \cdot 10^{34}\; L_H \quad ,
\ee
the model is consistent with the Bekenstein bound but a light trapping region forms around the origin.

Providing consistency with the Bekenstein entropy bound might be the clearest indication to-date that there exists no more than a finite amount of matter.  Other successes such as the avoidance of infinite replication of each individual, the avoidance of infinite numbers of monsters in the multiverse and the avoidance of the measure problem of standard cosmology are, perhaps, more philosophical in nature as is the no-cosmic-light-trapping requirement.  Nevertheless, it
is a satisfying result of the entropy bound that the upper limit on the $R$ parameter found here assures that there is a negligible probability to produce two identical but unrelated human beings anywhere in the universe.

\section {Acknowledgements}
We thank Gary Goldstein, Ali Masoumi, and Ken Olum of Tufts University for helpful discussions.


\begin{thebibliography}{99} 
\bibitem{CG1} L. Clavelli and Gary Goldstein, Int. J. Mod. Phys. A28, 1350148 (2013). 
\bibitem{CG2} L. Clavelli and Gary Goldstein, Int. J. Mod. Phys. A30, 1550068 (2015). 
\bibitem{Bekenstein} J.D. Bekenstein, Phys. Rev. D7, 2333 (1973).\\
J.D. Bekenstein, Phys. Rev. D23, 287 (1981).\\
J.D. Bekenstein, Int. J. Theor. Phys. 28, 969 (1989).
\bibitem{MasoumiMathur} A. Masoumi and S.D. Mathur, Phys. Rev. D91, 8, 084058 (2015) ArXiv:1412.2618.
\bibitem{OlivePeacock} K.A. Olive and J.A. Peacock, Review of Big-Bang Cosmology, in Review of Particle Properties, J. Beringer et al. (PDG)
Phys. Rev. D86, 010001 (2012).
\bibitem{tHooft} G. 't Hooft ''Dimensional reduction in quantum gravity'' in ''Salam-festschrift'' Eds. A. Aly,  J. Ellis,  and S. Randjbar--Daemi, World Scientific Press (1993)  ArXiv:gr-qc/9310026).
\bibitem{Susskind} L. Susskind (1995) J Math. Phys.  36, 6377 (1995)  ArXiv:hep-th/9409089.
\bibitem{Bousso} R. Bousso, J High Ener. Phys. 9907:004 (1999), 
ArXiv:hep-th/9905177.
\bibitem{BrusteinVeneziano} R. Brustein and G. Veneziano, Phys. Rev. Letters  84, 5695 (2000), ArXiv:hep-th/9912055.
\bibitem{Frampton} P. Frampton,  Int. J. Mod. Phys. A30, 21, 1550129 (2015).
\bibitem{Pietronero} L. Pietronero, Physica A (144): 257 (1987).
\bibitem{deVega} H.J. de Vega, N. Sanchez, and F. Combes,  Ap. J. 500: 8 (1998), Astro-ph/9801224.
\bibitem{Hogg} D.W. Hogg et al., Ap. J. 624: 54 (2004), Astro-ph/0411197.
\bibitem{Tegmark} M. Tegmark, Ap.J. 606 (2): 702 (2004), Astro-ph/0310725.
\bibitem{Beckwith} S. Beckwith et al. Astro-ph/0607632, Astron. J. 132, 1729 (2006). 
\end{thebibliography}
\end{document}